\documentclass[dvips]{article}
\usepackage{lscape,epsfig}
\usepackage{amssymb}
\usepackage{amsmath}
\usepackage{rotating}
\DeclareSymbolFont{ppa}{OT1}{ppl}{m}{it}
\DeclareMathSymbol{\vv}{\mathalpha}{ppa}{'166}

\begin{document}   

\newcommand{\dd}{\,{\rm d}}
\newcommand{\ie}{{\it i.e.},\,}
\newcommand{\etal}{{\it et al.\ }}
\newcommand{\eg}{{\it e.g.},\,}
\newcommand{\cf}{{\it cf.\ }}
\newcommand{\vs}{{\it vs.\ }}
\newcommand{\zdot}{\makebox[0pt][l]{.}}   
\newcommand{\up}[1]{\ifmmode^{\rm #1}\else$^{\rm #1}$\fi}
\newcommand{\dn}[1]{\ifmmode_{\rm #1}\else$_{\rm #1}$\fi}
\newcommand{\upd}{\up{d}}
\newcommand{\uph}{\up{h}}
\newcommand{\upm}{\up{m}}
\newcommand{\ups}{\up{s}}
\newcommand{\arcd}{\ifmmode^{\circ}\else$^{\circ}$\fi}   
\newcommand{\arcm}{\ifmmode{'}\else$'$\fi}
\newcommand{\arcs}{\ifmmode{''}\else$''$\fi}
\newcommand{\MS}{{\rm M}\ifmmode_{\odot}\else$_{\odot}$\fi}
\newcommand{\RS}{{\rm R}\ifmmode_{\odot}\else$_{\odot}$\fi}
\newcommand{\LS}{{\rm L}\ifmmode_{\odot}\else$_{\odot}$\fi}

\newcommand{\Abstract}[2]{{\footnotesize\begin{center}ABSTRACT\end{center}
\vspace{1mm}\par#1\par
\noindent
{~}{\it #2}}}

\newcommand{\TabCap}[2]{\begin{center}\parbox[t]{#1}{\begin{center}
  \small {\spaceskip 2pt plus 1pt minus 1pt T a b l e}
  \refstepcounter{table}\thetable \\[2mm]
  \footnotesize #2 \end{center}}\end{center}}

\newcommand{\TableSep}[2]{\begin{table}[p]\vspace{#1} 
\TabCap{#2}\end{table}}

\newcommand{\FigCap}[1]{\footnotesize\par\noindent Fig.\  %
  \refstepcounter{figure}\thefigure. #1\par}

\newcommand{\TableFont}{\footnotesize}
\newcommand{\TableFontIt}{\ttit}
\newcommand{\SetTableFont}[1]{\renewcommand{\TableFont}{#1}}

\newcommand{\MakeTable}[4]{\begin{table}[htb]\TabCap{#2}{#3}
  \begin{center} \TableFont \begin{tabular}{#1} #4
  \end{tabular}\end{center}\end{table}}
  
\newcommand{\MakeTableSep}[4]{\begin{table}[p]\TabCap{#2}{#3}
  \begin{center} \TableFont \begin{tabular}{#1} #4
  \end{tabular}\end{center}\end{table}}

\newenvironment{references}%
{
\footnotesize \frenchspacing
\renewcommand{\thesection}{}
\renewcommand{\in}{{\rm in }}
\renewcommand{\AA}{Astron.\ Astrophys.}
\newcommand{\AAS}{Astron.~Astrophys.~Suppl.~Ser.}
\newcommand{\ApJ}{Astrophys.\ J.}
\newcommand{\ApJS}{Astrophys.\ J.~Suppl.~Ser.}
\newcommand{\ApJL}{Astrophys.\ J.~Letters}
\newcommand{\AJ}{Astron.\ J.}
\newcommand{\IBVS}{IBVS}
\newcommand{\PASP}{P.A.S.P.}
\newcommand{\Acta}{Acta Astron.}
\newcommand{\MNRAS}{MNRAS}
\renewcommand{\and}{{\rm and }}
\section{{\rm REFERENCES}}
\sloppy \hyphenpenalty10000
\begin{list}{}{\leftmargin1cm\listparindent-1cm
\itemindent\listparindent\parsep0pt\itemsep0pt}}%
{\end{list}\vspace{2mm}}

\def\TYLDA{~}
\newlength{\DW}
\settowidth{\DW}{0}
\newcommand{\dw}{\hspace{\DW}}
  
\newcommand{\refitem}[5]{\item[]{#1} #2%
\def\REFARG{#3}\ifx\REFARG\TYLDA\else, {\it#3}\fi
\def\REFARG{#4}\ifx\REFARG\TYLDA\else, {\bf#4}\fi 
\def\REFARG{#5}\ifx\REFARG\TYLDA\else, {#5}\fi.}

\newcommand{\Section}[1]{\section{#1}}
\newcommand{\Subsection}[1]{\subsection{#1}}
\newcommand{\Acknow}[1]{\par\vspace{5mm}{\bf Acknowledgements.} #1}
\pagestyle{myheadings}

\newfont{\bb}{ptmbi8t at 12pt}
\newcommand{\xrule}{\rule{0pt}{2.5ex}}
\newcommand{\xxrule}{\rule[-1.8ex]{0pt}{4.5ex}}
\def\thefootnote{\fnsymbol{footnote}}
\begin{center}

{\Large\bf
Variable Stars in the
Globular Cluster NGC~6752\footnote{
This paper includes data gathered with the 6.5 meter Magellan Telescopes
located at Las Campanas Observatory, Chile.}
}

\vskip1cm
{
\large J. Kaluzny$^{1}$ and I.~B. Thompson$^{2}$}
\vskip3mm
{
       $1$ Nicolaus Copernicus Astronomical Center,\\
        ul. Bartycka 18, 00-716 Warsaw, Poland\\
        e-mail: jka@camk.edu.pl\\

      $^{2}$Carnegie Institution of Washington, 813 Santa Barbara Street,
      Pasadena, CA 91101, USA \\
      e-mail: ian@ociw.edu\\
}
\end{center}

\Abstract{
We report time-series photometry for 16 variable stars located
in the central part of the globular cluster NGC~6752. 
The sample includes 13 newly identified objects. The precision of our
differential photometry ranges from 1~mmag at $V=14.0$ 
to 10~mmag at $V=18.0$. We detected 4 low amplitude
variables located on the extended horizontal branch (EHB) of the cluster. 
They are candidate binary stars harboring sdB subdwarfs.
A candidate  degenerate binary was detected about 2 mag below
the faint end of the EHB.
The star is blue and its light curve is modulated with a period
of 0.47~d. We argue that some of the identified variable red/blue stragglers 
are ellipsoidal binaries harboring degenerate stars. They have
low amplitude sine-like light curves and periods from a few hours 
to a few days.
Spectroscopic observations of such objects may lead to  the
detection of detached inactive binaries harboring stellar mass black holes or
neutron stars. No binaries of 
this kind are known so far in globular clusters although their 
existence is expected based on the common occurrence of accreting LMXBs
and millisecond pulsars. An eclipsing SB1 type binary was identified
on the upper main sequence of the cluster. We detected
variability of optical counterparts to two X-ray sources located in the
core region of NGC~6752. The already known cataclysmic variable
B1=CX4 experienced a dwarf nova type outburst. The light curve of an optical 
counterpart to the X-ray source CX19 exhibited modulation with a 
period  0.113~d. The same periodicy was detected in the HST-ACS data.
The variable is located on the upper main sequence of the cluster. It is 
an excellent candidate for a close degenerate binary observed in quiescence.    
}

{\bf Key words:} {\it
stars: dwarf novae - novae, cataclysmic variables -- globular
clusters: individual: NGC 6752 -- stars: horizontal-branch --
binaries: eclipsing}


\section{Introduction} \label{sect:intro}
NGC~6752 is a nearby globular cluster whose low reddening and
relatively high Galactic latitude 
($(m-M)_{V}=13.02$, $E(B-V)=0.04$, $b=-25.6$~deg; Harris 1996)
make it an attractive target for detailed studies.  
Optical counterparts for 12 out of 19 faint X-ray sources 
detected with $Chandra$ were reported by Pooley et al. (2002).
This sample included ten likely cataclysmic variables (CVs) and
1-3 RS~CVn or BY Dra stars. Two of the candidate CVs were detected 
and studied earlier by Bailyn et al. (1996) based on
HST/WFPC2 data.
So far there have been no reports of dwarf novae type outbursts in the field
of NGC~6752.
The cluster is known to host 
five millisecond pulsars (D'Amico et al. 2002), one of which has  an
optical counterpart (Ferraro et al. 2003; Bassa et al. 2003).
A wide-field CCD based survey of NGC~6752 conducted by 
Thomson et al. (1999) led 
to the detection of eleven photometric variables, seven of which were classified
as contact binaries, and three as SX~Phe stars. The issue of membership status
of these variables remains open. In particular,  Rucinski (2000) argues that  
the group of
contact binaries is dominated by field interlopers. 

The cluster has
a rich population of blue horizontal branch (BHB) and extreme horizontal 
branch (EHB) stars (Buonanno et al. 1986; Momany et al, 2002). 
Until recently not a single photometric variable was known
among the horizontal branch stars in NGC~6752. Catelan et al. (2008) 
reported possible detection of a BHB pulsator, while Kaluzny \& Thompson (2008)
detected four variables located on the BHB/EHB of the cluster. Only
one spectroscopic binary was found among a few dozen BHB/EHB stars
observed by Moni Bidin et al. (2008). This  result was 
unexpected, as close binaries are common among 
field sdB stars (Maxted et al. 2001; Napiwotzki et al. 2004). 

The photometric survey presented here was conducted 
as a part of the CASE project (Kaluzny et al. 2005). 
It is complementary to the wide-field study presented 
by Thompson et al. (1999). The new data, collected with a larger telescope 
at a finer xpatial scale, allowed a more detailed study of the central part of the cluster.
Since the pioneering work by Mateo et al. (1990) several globular clusters
have been surveyed with CCD photometry for faint variables, and 
in particular for main sequence binaries. These surveys, usually conducted
with 1-m class telescopes, are very incomplete in 
the cluster core regions, arising mainly from  
the saturation of stellar profiles of
densely packed bright stars. This saturation is hard to avoid
as exposures times have to be at least a few minutes long to assure 
sufficient S/N for the relatively faint main-sequence stars.  
The problem can be partially overcome by using larger telescopes 
with finer pixel scales producing images with a reduced
number of saturated stars.  
An even better solution is to use imaging capabilities of the $HST$, 
but so far only one cluster has been systematically surveyed for 
variability with this instrument 
(47 Tuc, Gilliland et al. 2000; Albrow et al. 2001). Results published for a few
other clusters observed with the $HST$ are based on 
fragmentary data. 

It is expected  that most of the eclipsing variables are
located in  the central parts of globular clusters. First, by definition,
roughly 50\% of all stars are observed within the half-light radius 
of a given cluster.
Second, dynamical 
considerations indicate that most of the binary stars migrate
toward  the core regions due to mass segregation (Stodolkiewicz 1986; 
Hut 1992). 
As we show below, our adopted observing strategy along with  
careful  data reduction has allowed a successful search
for variability in  a large sample of stars in the central part
of NGC~6752. Good quality photometry extends down to $V\approx 21$~mag 
and the light curves of the brightest stars show an rms 
of about  1~mmag. 

\section{Observations and data analysis}\label{sect:observations}

All images were taken at Las Campanas Observatory with the 2.5-m
du Pont telescope.  A field of $8.84\times 8.84$ arcmin$^{2}$ was 
observed with the TEK5 CCD camera at a scale of 0.259
arcsec/pixel. The cluster core was positioned roughly 1 arcmin north 
of the detector center to eliminate a nearby 
7th magnitude field star from the images. Observations were made on eight 
consecutive nights between 28 May and 
4 June 1998. Images were taken in two bands with average exposure times 35~s 
and 60~s for  the $V$ and $B$ filters, respectively. 
The readout time of the detector was 68~s. 
In total, 495 useful frames in $V$ and
287 frames in $B$ were collected. The field was monitored for
a total of about 30 hours. The median seeing was  1.09  and 1.15 arcsec  for 
the $V$ and $B$ bands, respectively.  
Sequences of 3-4 exposures in $V$ 
were interlaced with sequences of 2  exposures in $B$.
The images taken in a given sequence were combined, resulting
in 152 and 143 stacked frames for $V$ and $B$, respectively. The photometric
analysis was conducted separately for individual frames and for the stacked
frames. These stacked frames have a higher S/N ratio but 
have a poorer time resolution of about 10 minutes.
The search for variable stars was made with a modified
version of the ISIS image subtraction package 
(Allard \& Lupton 1998; Allard 2000), while
Daophot/Allstar codes (Stetson 1987) were used to extract lists of
point sources for the reference images. The observed field was divided into a
$3\times 3$ mosaic to reduce potential degradation of photometry 
caused by a spatially variable
point spread function. Light curves were extracted for 46905 and 
40078 point sources detected on the reference images 
for the $V$ and $B$ bands, respectively. The instrumental photometry was 
transformed to the standard $BV$ system using linear transformations 
based on 47 measurements of 16 stars from 4  Landolt (1992) fields.
Fig. 1 shows the rms of individual 
measurements versus average magnitude for light curves extracted from
single  as well as from stacked frames. 
Stars with $V<14.0$ are overexposed on the template images
and photometry of these objects
is unreliable. Photometry based on single frames has an accuracy of
about 2 mmag  for the brightest unsaturated stars,
decreasing to about 10 mmag at $V=18$. Light curves of
the brightest unsaturated  stars extracted from 
stacked frames show an rms of 1 mmag. 

The light curves were checked for variability with
AoV and AOVTRANS algorithms running in the TATRY 
program (Schwarzenberg-Czerny 1996, Schwarzenberg-Czerny \& Beaulieu 2006). 
The residual images produced with ISIS were also
searched for potential variables lacking counterparts in the list of
point sources detected with Daophot/Allstar in the template images.

A total of 15 variable stars were identified in our data. Their
equatorial coordinates along with  basic photometric
parameters are listed in Table 1. Equatorial coordinates of the 
variables were deteremined using 92 stars from the UCAC2 
catalog (Urban et al. 2004) which were identified on 
the $V$-band reference image.
Objects 7-12 were  reported by Thompson et al. (1999)
while objects 15-24 are new identifications \footnote{
Light curves of all variables discussed in this paper are available
from the CASE archive at http:/http://case.camk.edu.pl/}.
Variable number 25 is a
cataclysmic variable originally identified  by Bailyn et al. (1996)
based on $HST$ imaging.
Finding charts for the new variables are given in Fig. 2.   
Figure 3 shows the location of the detected variables (with the exception of
V25) on the 
cluster color-magnitude diagram (CMD).
The plotted positions correspond to magnitudes at maximum light and
average colors. Only 50\% of all stars with extracted $BV$ photometry 
are included in Fig. 3. Stars with formal errors of $V$ photometry
exceeding the median value of $\sigma_{\rm V}$ at a given magnitude level
have been  omitted. However, we have plotted all detected stars
with $B-V<0.3$ regardless of the quality of their photometry.

\begin{table}
\begin{center}
\caption{Equatorial coordinates and basic data for NGC~6752 variables
}
\begin{tabular}{|l|c|c|c|c|c|c|c|}
\hline
   ID &  RA(J2000)& Dec(J2000) & $V_{\rm MAX}$& $<B-V>$ & $\Delta V$ & P & Remarks \\
      &  [deg]    & [deg]      &              &         &            &[d]&  \\
\hline
V7 &287.72527 &-60.00345 &15.33 &   0.295 & 0.45 & 0.059051(2)& SX$^{a}$\\
V8 &287.79417 &-59.98151 &17.058&   0.365 & 0.36 & 0.314938(36)& EW$^{b}$\\
V9 &287.85929 &-60.02352 &15.018&   0.613 & 0.06 & 0.363602(53)& EW?\\
V12 &287.64263 &-59.94921 &16.292&  0.291 & 0.05 & 0.040901(2)& SX\\
V15 &287.79719 &-59.99593 &16.344& -0.146 & 0.02 & 0.6895(12)  & EHB$^{c}$\\
V16 &287.77680 &-59.98007 &16.53 & -0.163 & 0.04 & $>8$      & EHB\\
V17 &287.76690 &-59.98539 &15.27 &  0.111 & 0.02 & 3.291(12) & EHB\\
V18 &287.79165 &-60.03861 &20.74 & -0.294 & 0.16 & 0.4676(11)& EHB \\
V19 &287.80435 &-59.92156 &16.321&  1.072 & 0.12 & 6.18:   &  \\
V20 &287.71042 &-59.96032 &16.090&  0.742 & $>0.02$ & $>8$      & -\\
V21 &287.68723 &-60.00998 &16.477&  0.336 & 0.09 & 0.5514(15)& \\
V22 &287.62364 &-59.94002 &18.437&  0.769 & 0.13 & 1.76(10)& \\
V23 &287.63577 &-60.02607 &18.463&  0.537 & 0.2  &    -    & EA$^{d}$\\
V24 &287.62177 &-60.01576 &14.606& -0.054 & 0.01 & $>8$      & BHB, var?\\
V25 &287.71479 &-59.98371 &  -   &    -   &  -   &  -      & CX4, B1, CV$^{e}$\\
\hline
\end{tabular}
\end{center}
\vspace*{-0.3cm}
{\footnotesize Note: $^{a}$~SX~Phe type variable, $^{b}$~contact binary,
$^c$~extreme horizontal branch star, $^{d}$~detached eclipsing binary,
$^{e}$~cataclysmic variable B1 (Bailyn et al. 1996), optical 
counterpart to the X-ray source CX4 (Pooley et al. 2002).}
\end{table}

\section{Properties of Variables}\label{sect:vars}

\subsection{Horizontal Branch Stars} 
Our sample includes 239 stars with $B-V<0.22$ and $14<V<16.0$. These
are the BHB and EHB stars of NGC~6752. The light curves of these
stars were examined in detail on an individual basis.
Seventy-two  objects with $16<V<18$ and $B-V<0.22$ are
candidate B-type subdwarf (sdB) stars. 

Two types of pulsating stars 
are known among  the field sdB stars. The p-mode pulsators (sdBV or EC~14026 
type stars) have periods of 100-200~s and 
amplitudes ranging from a few millimagnitudes up to 0.25-mag. 
The g-mode pulsators (PG1718--426 type stars) have periods ranging from 
20 minutes to 3 hours with amplitudes not exceeding 5~mmag.    
The sdB stars also occur in close binaries with orbital periods  
ranging from a few hours to several days 
(Morales-Rueda et al. 2003) and amplitudes of their light curves ranging 
from a few mmag to a few tenths of a magnitude. 
Most often the dominant mechanism of photometric variability of sdB binaries 
is the reflection effect.  

The time resolution of photometry measured from the individual frames 
is sufficient to detect coherent variations with periods of the 
order of 100~s. As can be seen in Fig. 1, the precision of photometry 
at $V\approx 18$ is about 10 mmag, improving to 4 mmag at $V=16.0$.
With light curves containing 495 data points we should be able to 
detect sdB stars pulsating with amplitudes as small as 1-2 mmag.
Not a single candidate for a pulsating sdB star was detected in our sample.
We have identified, however, four variables located in the BHB/EHB region
on the CMD of the cluster. A fifth blue variable is 
located on the faint extension of the EHB. Time domain $V$ band light curves
of these five variables are presented in Fig. 4. 
The $B$ band light curves are similar to those observed in $V$.

Stars V24 and V16 showed some systematic changes of luminosity
during our observations.
However, their possible periods exceed the eight day
interval spanned by our data.
The light curves of the remaining three variables can be phased with
preliminary periods, as  shown in Fig. 5.

For star V17 we have adopted a  period of $P=3.1$~d, with $P=6.2$~d
also a possibility. The location of this variable
on the cluster CMD (see Fig. 3) indicates that it may be a binary composed
of a hot EHB star and a red companion. In such a case the observed 
variability could be due to the reflection and/or ellipsoidality effect.  

The light curve of V15 was phased with $P_{1}=1.318$ (representing
the highest peak in a power spectrum of the V15 light curve), but periods  
$P_{2}=0.687$~d or $P_{3}=0.406$~d are also possible.  The variable is a good
candidate for binary sdB star.

The light curve of V18 was phased with a period of 0.4673~d. 
Anexamination of nightly light curves indicates that the
detected variability is very likely real although not strictly regular.
This claim is further supported by the photometry in the $B$ band.  
The phased $B$ light curve of V18  is presented in Fig. 6. 
It has a similar shape but a slightly larger amplitude than the $V$ curve.
We have used  images of NGC~6752 collected in 2007
and 2008 season to look for  possible long term variability of V18.
There was no evidence for any change of the average $V$ luminosity 
exceeding 0.1 mag with respect to the 1998 season.
As a result it seems unlikely that V18 is a  dwarf nova. 
The location on the cluster CMD indicates that it 
may be a degenerate binary hosting a low-mass helium white dwarf.
A binary of this type and at a similar location on the cluster CMD 
was found in the globular M4 (O'Toole et al. 2006).
The  suggested  orbital  period of V18 of 0.46~d
falls within the range occupied by low mass degenerate 
binaries (Downes et al. 2001).


\subsection{Blue Stragglers}\label{sect:BS}

Four of the detected variables are candidate blue stragglers.
Objects V7 and V12 are SX~Phe stars (Thompson et al. 1999). 
Our new photometry indicates that they are multi-modal 
pulsators. In the light curve of V7 we detected three periodicities:
$P_{1}=0.059051(2)$~d, $P_{2}=0.06152(2)$~d,
$P_{3}=0.040519(1)$~d. A fourth possible period  is $P{4}=0.05691(4)$~d or
$P{4}=0.06027(4)$~d -- the power spectrum of the pre-whitened light curve has
equal height peaks at these two periods. In the case of V12  
we detected two periods: 
$P_{1}=0.040901(2)$~d and $P_{2}=0.039509(7)$~d. The light curves of V7 and V12
phased with periods $P_{1}$ are shown in Fig. 7.

The blue straggler V21 is a likely $\gamma $~Dor type pulsator. Its 
time-domain light curve is presented in Fig. 8. It shows periodic modulation 
with a variable amplitude. The light curve phased with the dominant
period of 0.5513~d is presented in Fig. 9. The variable becomes bluer at 
the maximum light with a total range of $B-V$ of about 0.05~mag and an average
color $<B-V>=0.335$. The average magnitude is  
$<V>=16.54$. Assuming cluster membership  for V21 implies
an absolute magnitude  $M_{\rm V}=3.41$ and unreddened color 
$(B-V)_{0}=0.295$.
\footnote{Following Harris (1996) we adopt for the cluster
$(m-M)_{\rm V}=13.13$ and $E(B-V)=0.04$.}
The observed  characteristics of V21 are fully consistent with 
those of a $\gamma $~Dor star (Henry, Fekel \&  Henry 2005)
belonging to NGC~6752.
A spectroscopic confirmation of pulsations of
V21 would be a challenging task, as radial velocity amplitudes
of $\gamma $~Dor stars are on the level of a few km/s.


The last of the variable blue stragglers, V8, was originally
detected by Thompson et al. (1999). It has a classical W~UMa type 
light curve with period of 0.31~d and a total primary eclipse.
A detailed analysis of this star
will be presented separately.  V8 is a semidetached 
system, which is  unexpected given its short orbital period 
and the shape of the light curve.

\subsection{Other variables}\label{sect:other}

Variables V19 and V20 are located to the right of the subgiant branch
on the cluster CMD. Their light curves are shown in Fig. 10.
The variability of V19  seems to be periodic with $P\approx 6$~d ( 
$P\approx 1.2$~d is also possible). The $B-V$ color varies by 
about 0.018 mag, the star is bluer at maximum light. 
V20 shows a  systematic decline in luminosity during the observing 
run. Over eight nights the $V$ magnitude decreased by about 0.018 mag, 
while  the $B-V$ color remained constant. 

The time domain light curve of V22 is shown in Fig. 10. The variable is
located about 1 magnitude above the upper main-sequence of the cluster.
Its light curve can be phased with sevaral periods, of which $P=1.872$~d
corresponds to the highest peak in the power spectrum.  
Problems with aliasing and insufficient time coverage do not allow us
to derive the period of variability of V22 with confidence.


The location of variable V9 on the cluster CMD makes it a candidate 
yellow straggler. It shows variability  with a period of 0.364~d.
The phased light curve is shown in Fig. 11. The period and shape of the
light curve suggest that V9 is a low inclination W UMa type binary,
and is a field star not related to the cluster.
Nevertheless the nature and membership status of this relatively bright star
are worth of further study.

As can be seen in Fig. 12, variable V23 is an eclipsing binary. Only
one eclipse was observed. The star is located on
the main-sequence of the cluster. Such binaries are potentially valuable
as they can be used for accurate distance determination of the host cluster
(Paczy\'nski 1997). 
The observed eclipse of V23 has a flat 
bottom suggesting that the eclipse is total.
There was no detectable change of the $B-V$ color 
between maximum and minimum light implying that 
the eclipse is in fact a transient of a substantially cooler 
and smaller companion  in front of the primary. We suppose 
that the system has a large mass and luminosity ratio.
V23 was observed with the MIKE spectrograph on the Clay 6.5-m Magellan
Telescope on the nights of UT 04 August 2007 and UT 16 August 2007.
Analysis of the spectra with the IRAF FXCOR routine show that the star
is an SB1 binary with no evidence of a secondary component. The
measured heliocentric velocities were -23.23 +/- 0.19 km/sec at
HJD 2454316.5889 (mid exposure), and -40.32 +/- 0.20 km/sec at
HJD 2454328.6134.

The last object detected using our standard
procedure described in Sec. 2 is the known cataclysmic variable 
B1 (Bailyn et al. 1996). It  corresponds to  the X-ray source CX4
(Pooley et al. 2002) and  is located at an angular distance
of only 5~arcsec from  the cluster center. In Table 1 B1=CX4 is listed as 
V25. Several bright red stars are seen close to the variable on our
ground based images, causing  problems with the extraction of the 
$V$ band photometry.
However, we managed to extract $B$ band photometry. 
The variable underwent an outburst during 
our observations.  The light curve, presented in Fig. 13, is typical 
for ordinary U~Gem  type dwarf novae. V25 can be added to a short list 
of dwarf novae known in globular clusters (Pietrukowicz et al. 2008).

We have also looked  for variable objects at positions
corresponding to the remaining 18 X-ray sources listed in Pooley et al. (2002).
In the $B$ band, light curves could be extracted
for all locations but for the source  CX1. 
For the $V$ band some of examined locations were badly affected by 
nearby saturated stars.
A clear signature of variability was detected at the position 
of  CX19. A stellar object is seen at this location.
As can be seen in Fig. 14 it shows substantial night-to-night changes 
of the average luminosity. A closer examination reveals a 
sine-like modulation of the light curve with a period of about 0.11~d.
We measured the median luminosity of CX19 for each of the eight  
nights of data and offset the observations for each night
relative to the first night. The resultant 
$V$ light curve phased with a period of 0.11306~d is shown in Fig. 15. 
The variable is present on a series of HST ACS-WF images collected 
on 2004 September 19 between
9:13 and 17:49~UT (Proposal ID 10121; PI Bailyn). 
A total of 23 and 12 images were taken with 
F555W and F814W filters, respectively.   We  extracted profile photometry from these
frames using the Daophot/Allstar package. The photometry was transformed to the 
$VI_{C}$ system using the calibration provided by Sirianni et al. (2005).
Fig. 16 shows the ACS light curves of CX19  
phased with the period detected in our ground based data. The finding chart
for the optical counterpart of CX19 is shown in Fig. 17.
The variable is blended with  two close visual companions with
CX19 being the brightest, north-east, component of the blend. 
Components A and B are separated from CX19 by 0.12 and 0.21~arcsec,
respectively. Their magnitudes and colors are: $V_{A}=19.35$, $(V-I)_{A}=0.76$,
$V_{B}=21.11$, $(V-I)_{B}=1.27$. Components A and B could not be resolved
on the duPont images. However, we have taken them  into account
when transforming the ground-based $V$ light curve from 
differential counts into magnitudes.
Figure 18 shows a $V/(V-I)$ CMD extracted from the ACS data for a small 
region around CX19.
The variable is located very close to the upper main sequence of the cluster.
If it was a cluster member it would have an absolute visual magnitude
of $M_V\approx  4.9$. 

For the moment, it is difficult to assign CX19
to any class of variables with confidence.  
Its light curve resembles those of SU~UMa type CVs, but the 
period of 0.113~d is too long, and the observed luminosity is too high 
for such a classification. Moreover, SU UMa stars are much fainter
between outbursts than $M_V=4.9$. 
The value of the observed period falls within the range occupied by polars. 
However, the   X-ray luminosity of CX19 is much too 
low for an active magnetic CV. Assuming cluster membership, 
Pooley et al. (2002) obtained $L_{X}(0.5-2.5)~{\rm keV}=2.2\times 10^{30}$~erg/s
while active polars have  $L_{X}>10^{32}$~erg/s. 
It is possible that at the time of the $CHANDRA$ observation the variable was 
in a low state. However
the relatively red color observed at  $M_V\approx 4.9$ 
rules out the possibility that  CX19 is a polar since in their high states 
polars are intrinsically blue stars. We also note  that 
observations taken in 1998 with the du Pont telescope and in 2004 with ACS show the 
variable at a similar visual magnitude of $V\approx 18.0$. This is  
unexpected for an active CV.
We conclude that  the variable is not a CV. 
Instead we propose that it is a close binary hosting a neutron star 
or a black hole. The observed modulation of the optical light curve could 
be due to the ellipsoidal effect, in which case the orbital period would
be $2\times 0.113=0.226$~d. Additional low-amplitude  variability seen 
on the time scale of days can be attributed to fluctuations of 
the very low accretion rate  from the degenerate component 
to its compact companion. Such a mass transfer rate may also account
for the low X-ray luminosity. Despite being located close
to the cluster center, the variable seems to be accessible for ground 
based spectroscopy, and such radial velocity observations should reveal the nature of CX19.


\section{Summary and Discussion}\label{sect:summary}

We have obtained time series $BV$ photometry for about
40 thousand 
stars from the central area of the globular cluster NGC~6752. 
The observing strategy
and careful reduction of the data resulted in a photometric
precision  of 4  mmag
stars with $V<16.0$  and 1-2 mmag at $V<15$, rising to 
 0.01 mag at $V\approx 18.5$.
The  sample included 72 hot subdwarf candidates. 
No pulsation variability was detected for any of them.
We have detected, however, four low-amplitude variables 
which may be binary EHB/BHB stars. 
Spectroscopic follow up is needed to reveal their actual nature.
Preliminary periods were established for two of them.
A faint blue variable with $V\approx 20.7$ was also detected. Its light
curve is likely periodic with $P\approx 0.47$~d. The variable is
a good candidate for a degenerate binary belonging to the cluster.
One of four periodically variable blue stragglers is
a likely $\gamma$~Dor star. If confirmed, it would be the first 
variable of this type detected in globular clusters.

We detected  three variable yellow/red stragglers with $15.0<V<16.3$. 
Two of them  show periodic
sine-like modulation of their light curves. 
Another periodic variable was detected 
at $V=18.4$ on the right side of the upper-main sequence of the cluster. 
We propose that their light-curve
modulation is related to binarity, and  that they   
are good candidates for ellipsoidal variables hosting degenerate components. 
Our hypothesis is based on two main arguments.
First, it is known that field X-ray novae (binaries hosting stellar mass
black holes) spend most of their time
in quiescence, showing only an ellipsoidal variability
(Remillard \& McClintock 2006).
Second, the known optical counterparts to millisecond pulsars
located in GCs usually occupy positions either to the red
or to the blue of the main sequence on the H-R diagram (Ferraro
et al. 2001, 2003). As GCs contain rich population of active
X-ray binaries, one may expect that they also harbor a large number
of non-accreting, degenerate binaries. Unambiguous detection of non-accreting
degenerate binaries in GCs would open a new, exciting field of research,
and may lead to the detection of the first known stellar-mass
black holes in GCs. The list of possible quiescent degenerate 
binaries in GCs is limited to a few faint X-ray sources of unknown nature 
(Heinke et al. 2003). The binary nature of red/yellow stragglers 
from NGC~6752 can be checked by obtaining a few medium resolution spectra
per object. These spectra  could be  used to estimate the 
mass function for confirmed binaries, further constraining the nature of 
possible compact components.

We have detected a likely dwarf nova type outburst of one
of the CV candidates located in the  cluster core
region. Periodic variability was detected for the optical 
counterpart of the faint X-ray source CX19 located in the core region
of the cluster. The star is located very close to 
the upper main sequence of the cluster, and is unlikely to
be an ordinary cataclysmic variable.  We propose that the object it is 
an  excellent candidate for a close degenerate binary caught 
in  quiescence. Despite being located in the crowded area the 
variable is accessible for ground based spectroscopy. 




   

\Acknow{

Research of JK is supported by the grant MISTRZ 
from the Foundation for the Polish Science and by the grant 
N N203 379936 from the Ministry of Science and Higher Education.
I.B.T. acknowledges the support of NSF grant AST-0507325.
Based on observations made with the NASA/ESA Hubble Space Telescope,
and obtained from the Hubble Legacy Archive, which is a collaboration
between the Space Telescope Science Institute (STScI/NASA), the Space
Telescope European Coordinating Facility (ST-ECF/ESA) and the
Canadian Astronomy Data Centre (CADC/NRC/CSA).
}

\newpage

\begin{figure}[htb]
\centerline{\includegraphics[width=120mm]{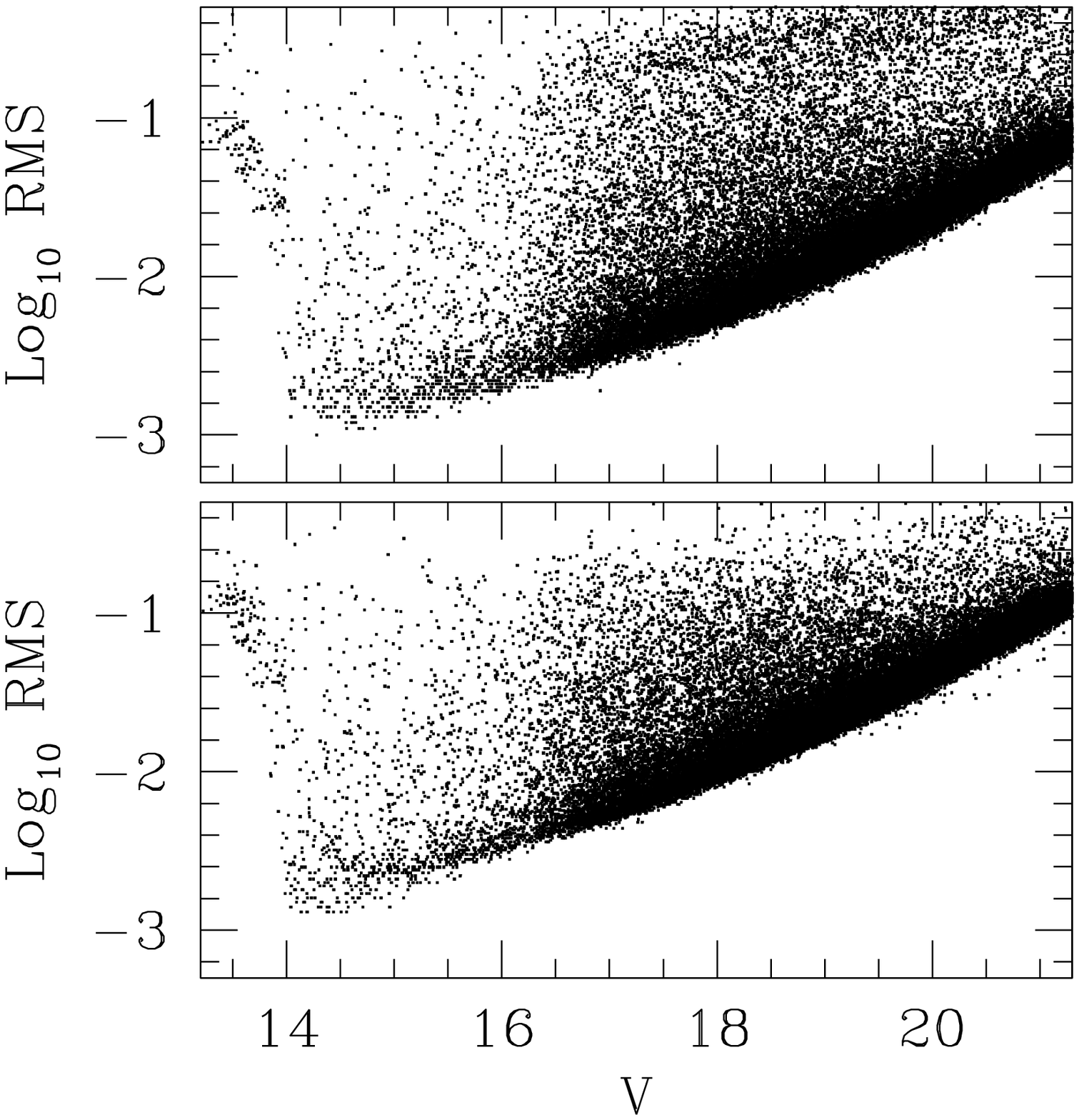}}
\caption{\small Standard deviation vs. average $V$ magnitude for
light curves based on individual $V$ frames (bottom panel) and for
light curves based on stacked $V$ frames (upper panel).}
\end{figure}

\begin{figure}[htb]
\centerline{\includegraphics[width=120mm]{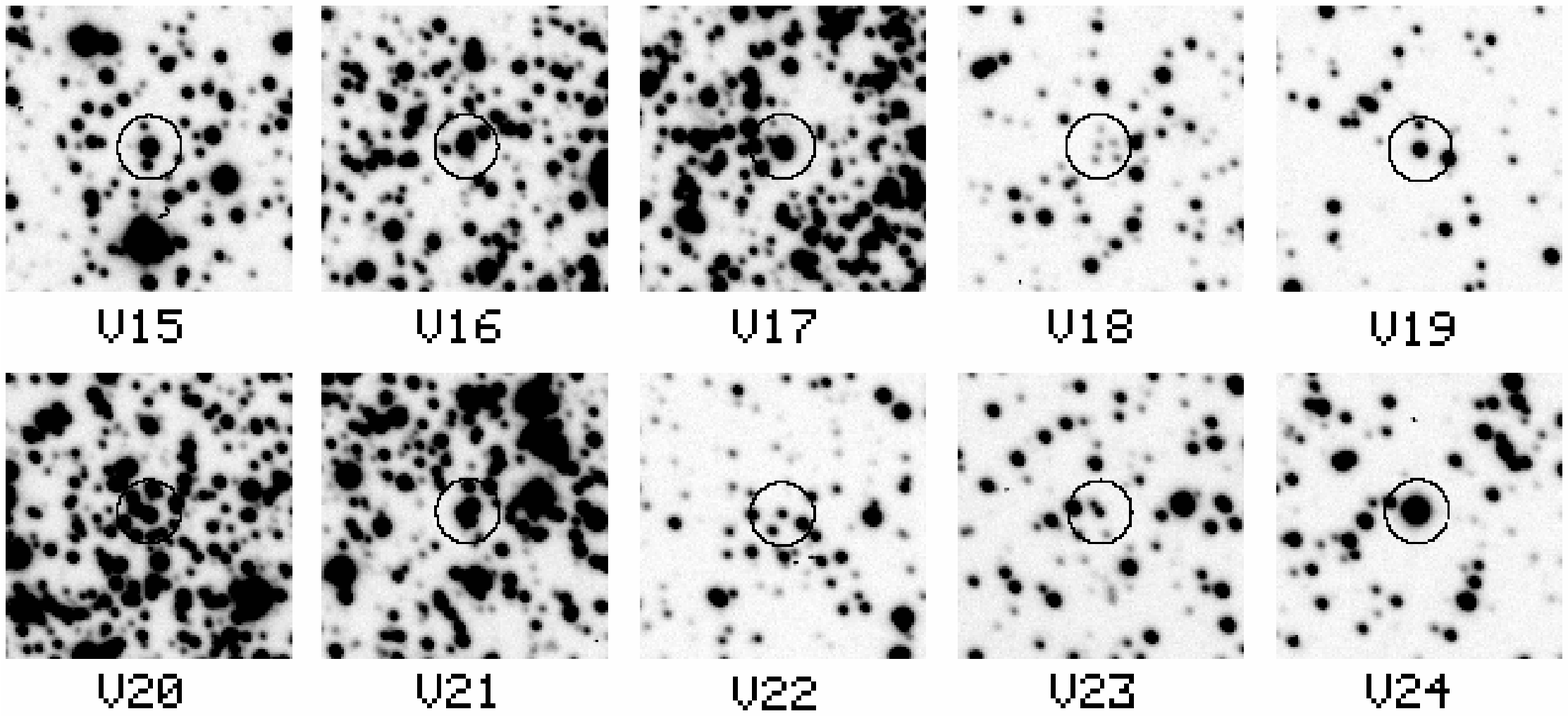}}
\caption{\small Finding charts for variables V15-24.
Each chart is 30 arcsec on a side: north is up and east to the left.}
\end{figure}

\begin{figure}[htb]
\centerline{\includegraphics[width=120mm]{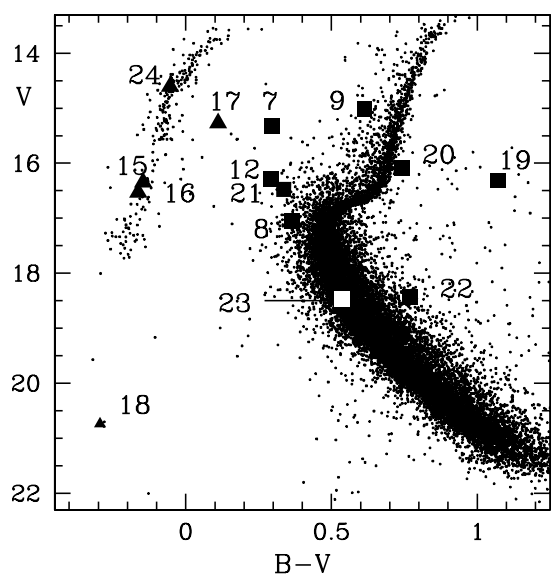}}
\caption{\small CMD for NGC~6752, with positions of the variables marked.}
\end{figure}

\begin{figure}[htb]
\centerline{\includegraphics[width=120mm]{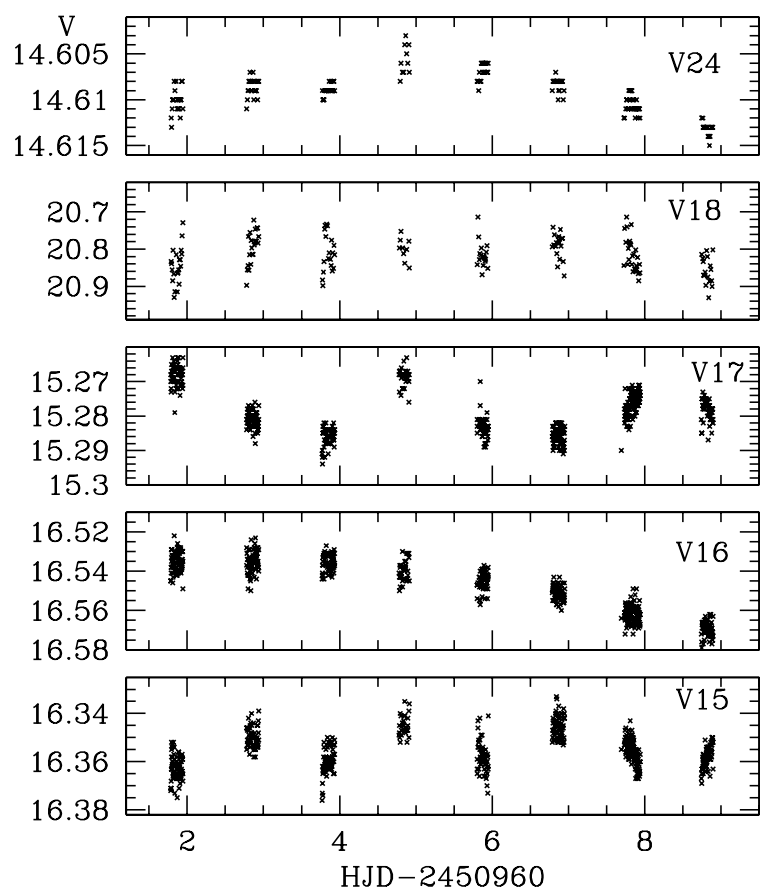}}
\caption{\small Light curves of variables from the BHB/EHB of NGC 6752.}
\end{figure}

\begin{figure}[htb]
\centerline{\includegraphics[width=120mm]{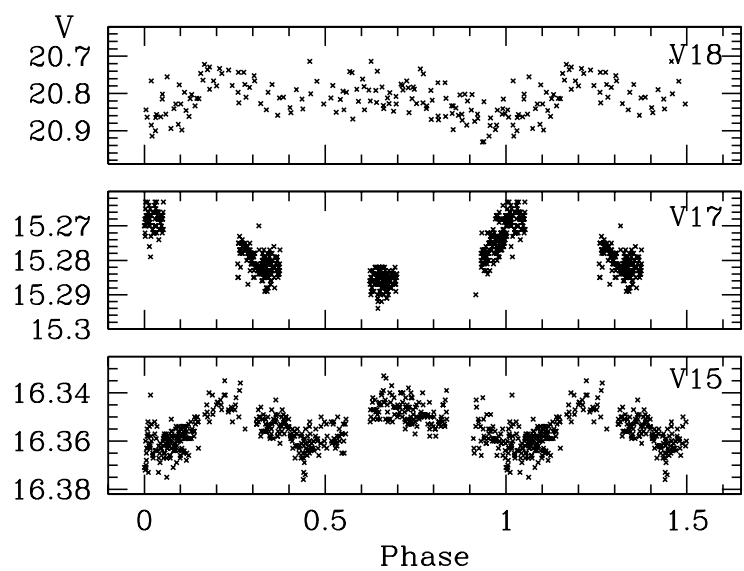}}
\caption{\small Phased $V$ light curves of the variables V15, V17 and  V18.}
\end{figure}

\begin{figure}[htb]
\centerline{\includegraphics[width=120mm]{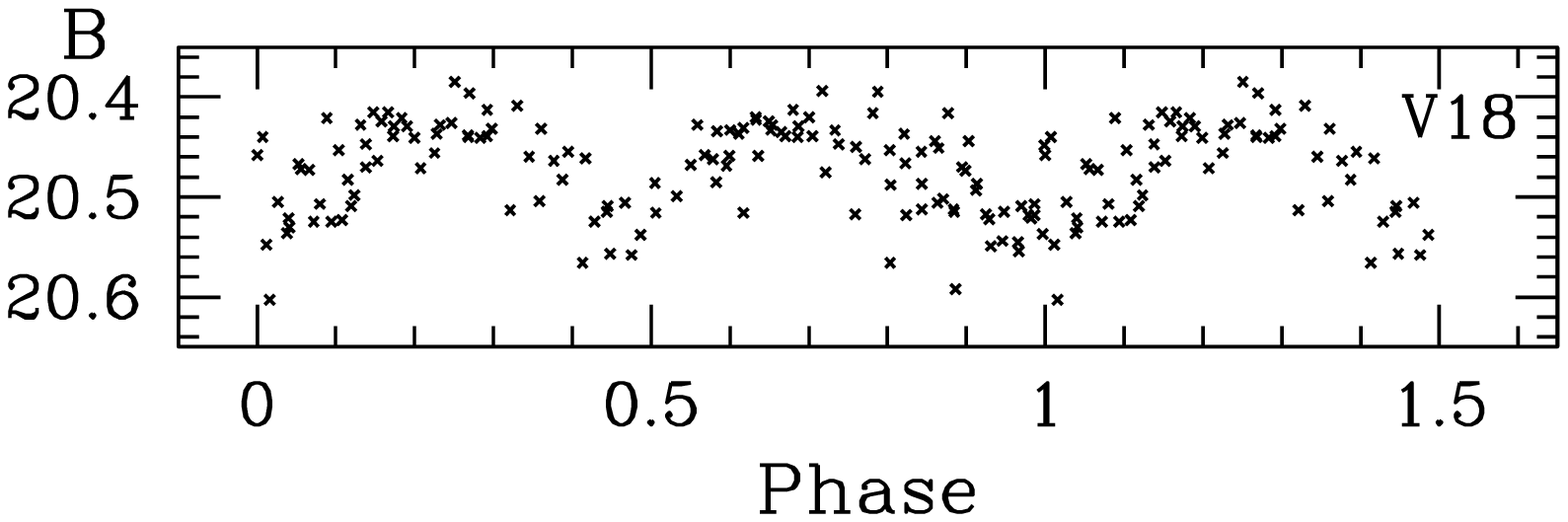}}
\caption{\small Phased $B$ light curve of the variable  V18.}
\end{figure}

\begin{figure}[htb]
\centerline{\includegraphics[width=120mm]{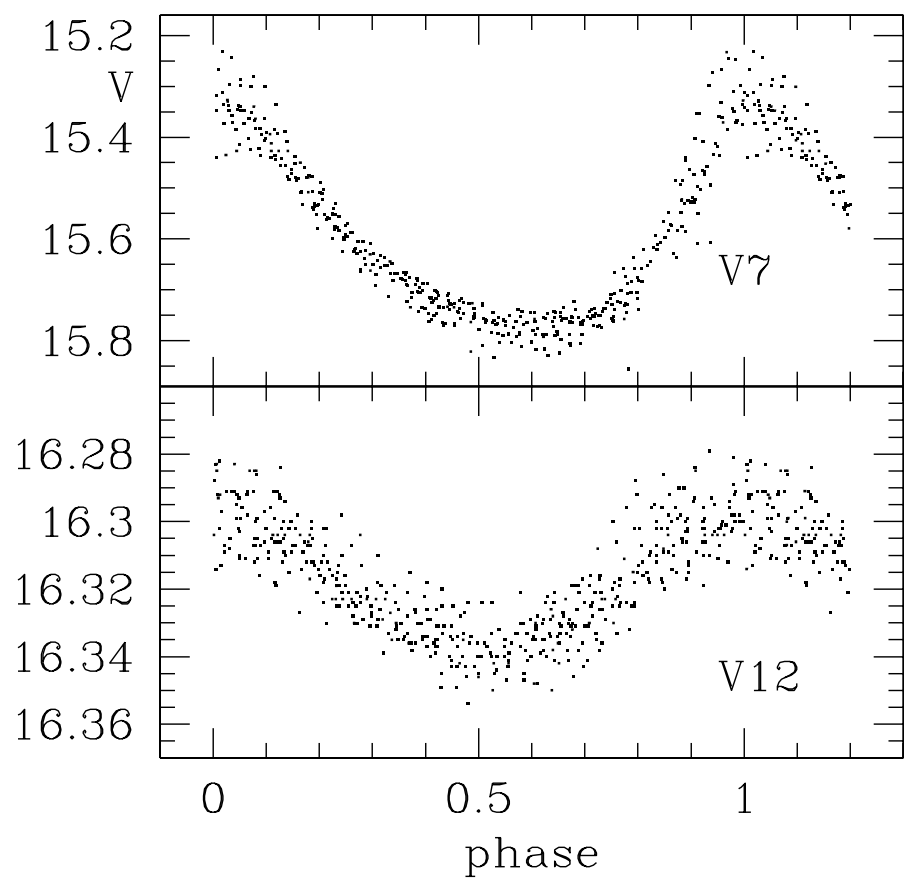}}
\caption{\small Phased $V$ light curves of SX~Phe stars V7 and V12.}
\end{figure}

\begin{figure}[htb]
\centerline{\includegraphics[width=120mm]{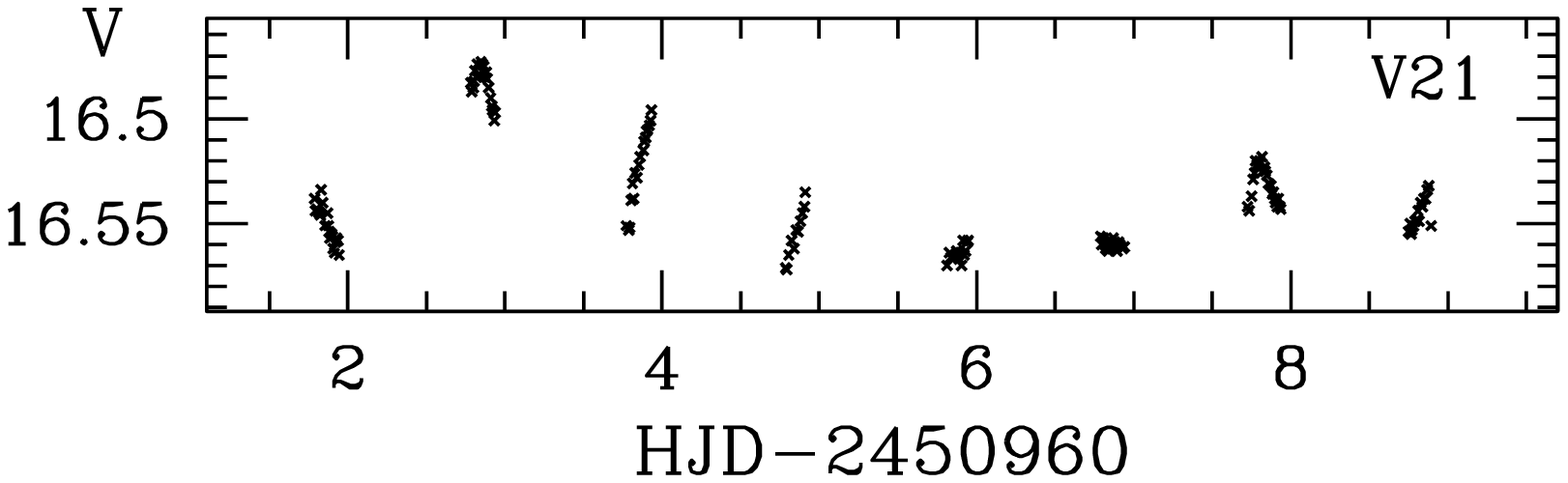}}
\caption{\small Light curve of the variable V21.
}
\end{figure}

\begin{figure}[htb]
\centerline{\includegraphics[width=120mm]{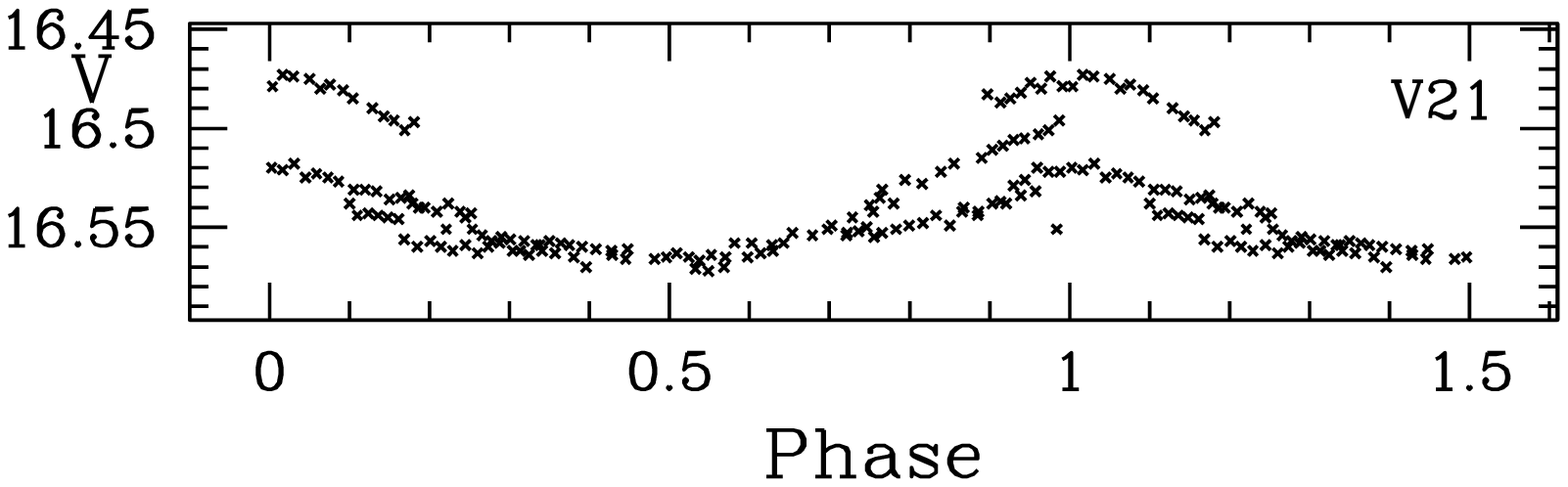}}
\caption{\small Phased $V$ light curve of the variable V21.
}
\end{figure}

\begin{figure}[htb]
\centerline{\includegraphics[width=120mm]{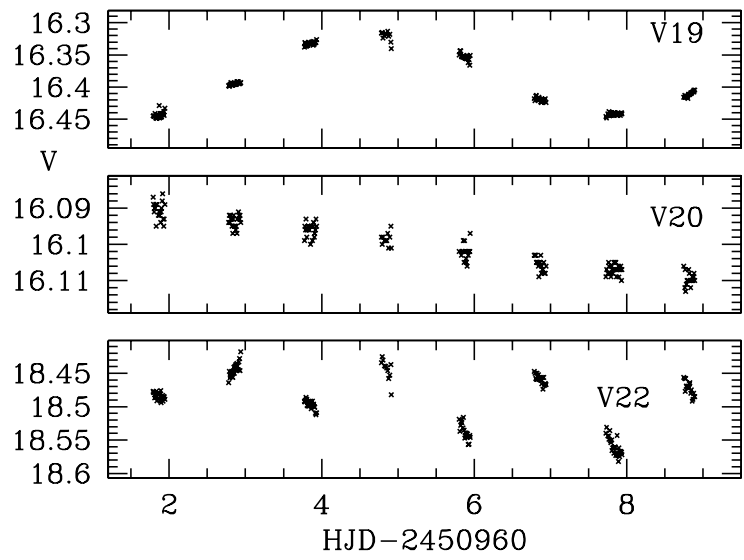}}
\caption{\small Light curves of the variables V19, V20 and V22.
}
\end{figure}

\begin{figure}[htb]
\centerline{\includegraphics[width=120mm]{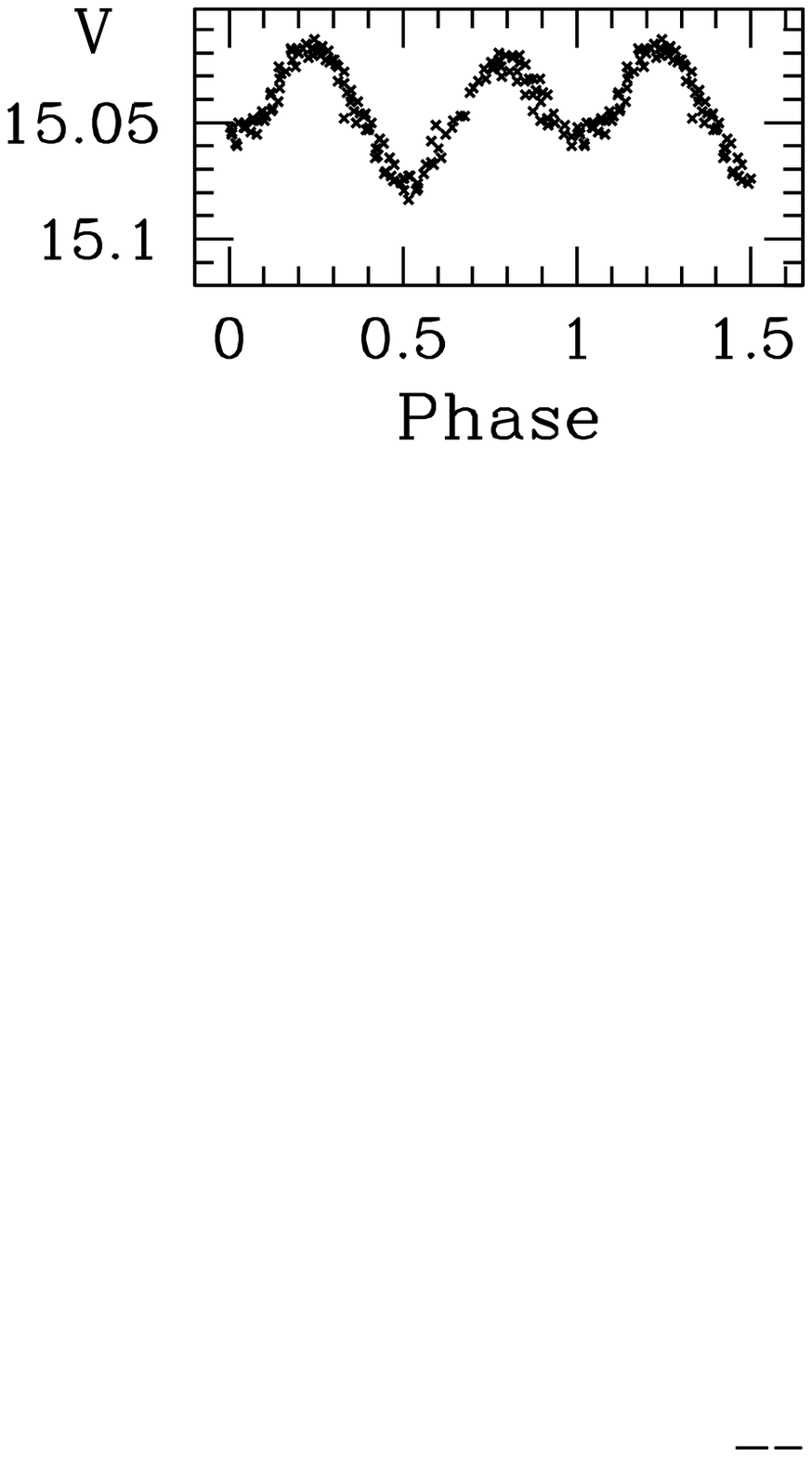}}
\caption{\small Phased $V$ light curve of the variable  V9.}
\end{figure}

\begin{figure}[htb]
\centerline{\includegraphics[width=120mm]{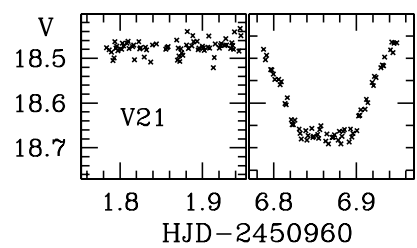}}
\caption{\small Light curve of the eclipsing binary V23.
}

\end{figure}

\begin{figure}[htb]
\centerline{\includegraphics[width=120mm]{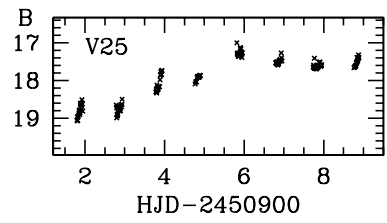}}
\caption{\small Light curve of cataclysmic variable B1=V25.
}
\end{figure}

\begin{figure}[htb]
\centerline{\includegraphics[width=120mm]{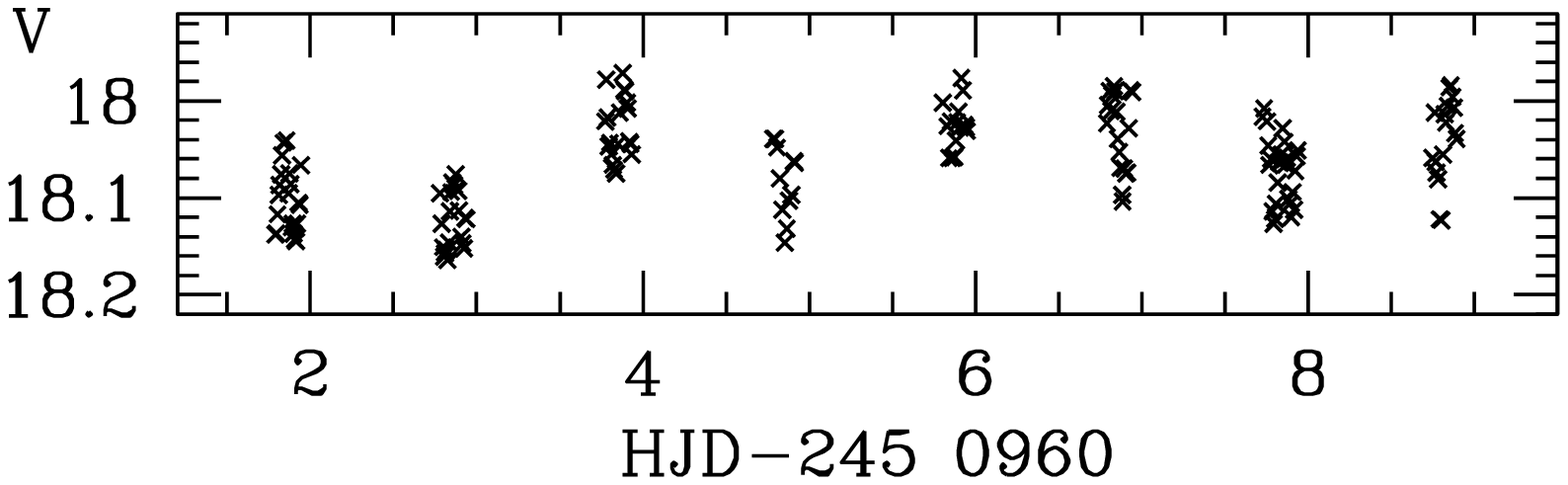}}
\caption{\small $V$ light curve of the
optical counterpart to the X-ray source CX19.
}
\end{figure}

\begin{figure}[htb]
\centerline{\includegraphics[width=120mm]{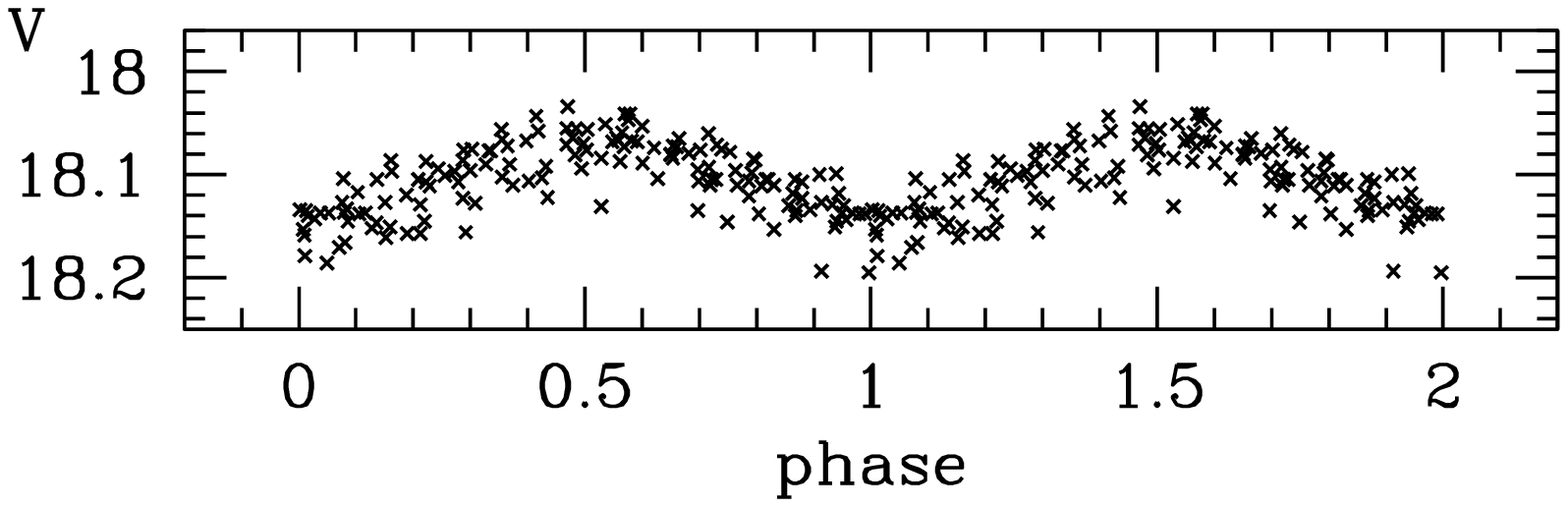}}
\caption{\small Phased $V$ light curve of the
optical counterpart to the X-ray source CX19.
}
\end{figure}

\begin{figure}[htb]
\centerline{\includegraphics[width=120mm]{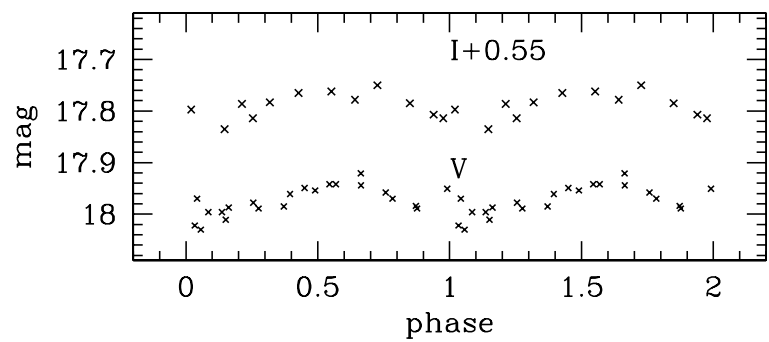}}
\caption{\small Phased $VI$ light curves of CX19 based on ACS data.
Note shift for the I band.
}
\end{figure}

\begin{figure}[htb]
\centerline{\includegraphics[width=120mm]{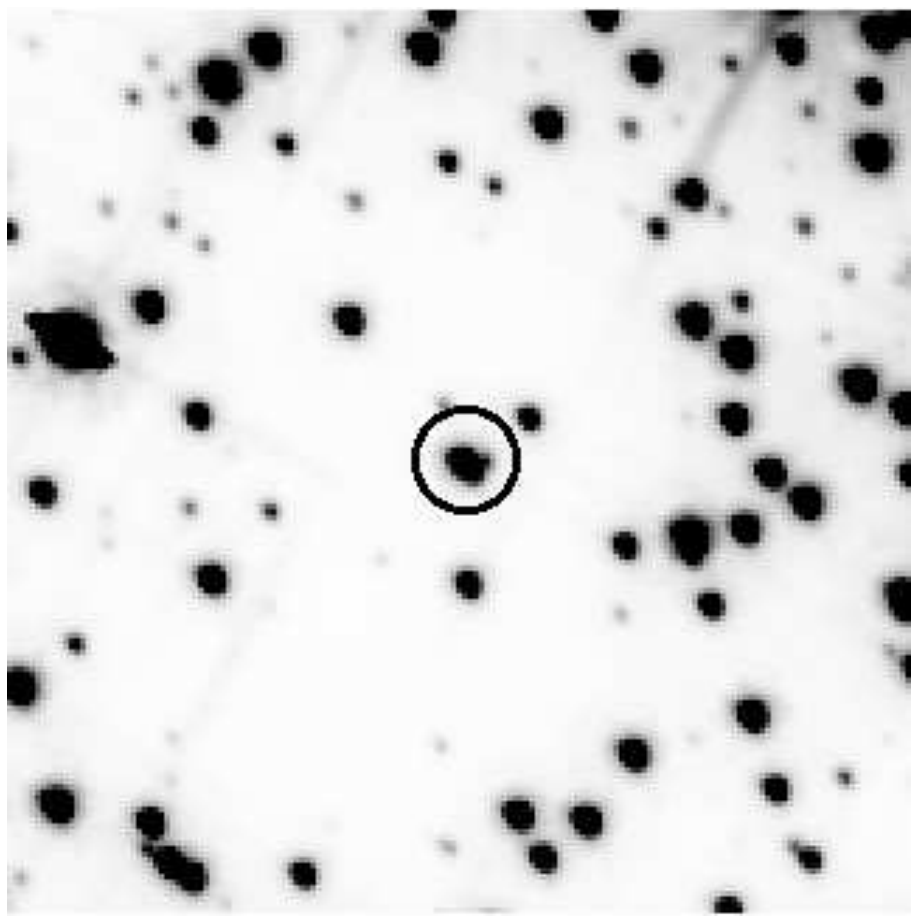}}
\caption{\small Finding chart for the variable CX19 based on the ACS image.
The chart is 4 arcsec
on a side with north up and east to the left. The variable is the N-E
component of the blend marked with the circle.
}
\end{figure}

\begin{figure}[htb]
\centerline{\includegraphics[width=120mm]{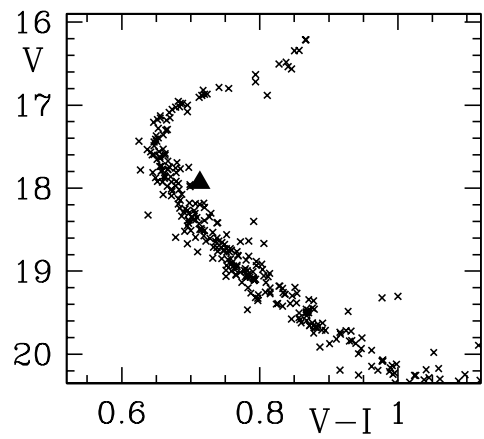}}
\caption{\small $V/V-I$ CMD of NGC~6752 with position of CX19 marked
with a triangle.
}
\end{figure}

\end{document}